# Intracellular regulatory control of gene expression process

Jafar Kazemi, Sadjad Ozgoli*

*Advanced Control Systems Laboratory, School of Electrical and Computer Engineering, Tarbiat Modares University, Tehran, Iran. (*Corresponding Author)*

**Abstract**

This paper presents an intracellular feedback control strategy, to regulate the gene expression process dynamics. For this purpose, two types of genetic circuits are designed in order to compare concentrations of the input transcription factor (the desired input) and the protein produced by the expression of target gene (process output). Genetic *Relay Switch* and *Shifted Subtractor* are proposed and employed to design *Discrete Comparator*. Also, another type of genetic comparator is presented named *Bistable Comparator*. These synthetic comparators are used to produce the required amount of the protein that activates the expression of the target gene. Numerical simulation results, demonstrate the effectiveness of the designed controllers.

*Keywords*: Synthetic Biology, Feedback Control, Relay Switch , Synthetic Comparator

## 1. Introduction

Considering the gene expression process as an input-output subsystem, design of genetic circuits by integration of these subsystems can be regarded as synthetic biology. For complex circuits, one may need to bring the concentration of a gene product to an arbitrary value. This target can be considered as a regulatory control problem. Control of gene expression process can lead to therapeutic applications such as restoring the correct secretion of insulin from pancreatic cells [1], or to create intracellular "disease fighters" [2]. Some other applications have been presented in [3].

Recently, various extracellular controllers have been employed to control of gene expression. For example, a PI-PWM controller has been designed to control a protein expression in population of yeast cells [4]. In [1] a simple switching control strategy is applied to in-vivo control of IRMA (In-vivo assessment of Reverse engineering and Modeling Approaches) network. A robust control method is used in [5] to regulate intracellular mRNA level via extracellular galactose in the GAL network as a well understood benchmark. In external control methods, firstly, output concentration is monitored by protein microscopy. The most common mechanism for this, is the coupling of green fluorescent protein (GFP) to the output protein and the use of image processing methods to estimate GFP concentration. After making a comparison between the measured output and the desired value, a control algorithm is implemented via an external computing device to determine the proper control action. Finally, the feedback control loop is completed by injecting the required value of the inducer to the cell.

From the practical point of view, the intracellular controllers can be more useful than the extracellular ones, in synthetic biology. Nevertheless, unlike the extracellular controllers, the intracellular ones are not investigated thoroughly in the literature. As one of the few works done in this field, an integral controller is implemented by a two-promoter genetic regulatory network for the purpose of disturbance rejection [6]. Even so, the design of a synthetic genetic controller for tracking control of gene expression remains as an unsolved problem.

Here we present, design of a synthetic genetic feedback controller for tracking control of gene expression process. For this purpose, two kinds of synthetic genetic circuits are designed to compare output genetic product with the desired input and produce a specific protein that activates the expression of the process gene. The design limitations are: the properties of the process input (the transcription factor that activates the process gene), the output (the protein produced by expression of the process gene) and the reference input (the transcription factor that is the reference for the process output). Changing the specification of these materials is impossible and the controller should be designed based on these limitations.

In this work, it is assumed that the transcription factor of the reference and process input act as activators. On the other hand, the output protein can be classified in the category of activator or repressor transcription factors. Therefore, two different synthetic comparators are designed: the *Discrete Comparator* to compare an activator transcription factor with a repressor transcription factor; and the *Bistable Comparator* for comparing two activator transcription factors. Synthetic *Shifted Subtractor* and *genetic Relay Switch* are designed for the first time and they are assembled in series to construct the *Discrete Comparator*. Also, the concept of synthetic bistable switch [7] is used to design a synthetic comparator named *Bistable Comparator*.

The rest of this paper is organized as follows: in section 2, the synthetic *Relay Switch* used in the design of *Discrete Comparator*, is presented. Synthetic comparators are designed in section 3. These comparators are used in feedback control of gene expression process in section 4. The numerical simulation results are presented in section 5. Finally, section 6 concludes the paper.

**2. Synthetic Relay Switch**

The *Relay Switch* is a device that allows its output to switch between two specified values base on its input. When the input is greater than the value of switch-on point, the relay will be on. On the contrary, if the input is less than switch-on point value, the relay will be off and the output will be equal to the minimum value.

**Proposition 1.** The genetic circuit depicted in Figure 1 acts as a genetic *Relay Switch*. The transcription factor $TF$ and the protein $P_2$ considered respectively as the input and the output of the *Relay Switch*. As shown in Figure 1, $TF$ and $P_2$ repress the expression of the first gene and $P_1$ represses the expression of the second gene.

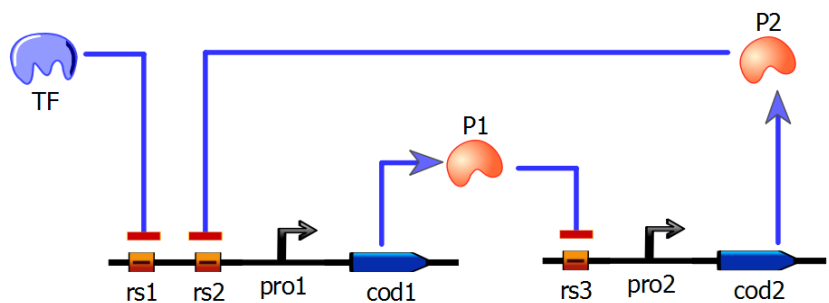

**Figure 1:** schematic of designed genetic *Relay Switch* in TinkerCell [8] software.

**Proof**:

In a high value of $TF$, the first gene will be repressed by $TF$ and concentration of $P_1$ will decrease. Then, by expression of the second gene the concentration of $P_2$ will increase. Increase of $P_2$ will cause a decrease in $P_1$ and this loop will go on to minimize $P_1$ and maximize $P_2$. Also, in a low value of $TF$, the concentration of $P_1$ will be increase for zero steady states. Increase of $P_1$ will repress expression of $P_2$ and the reduction of $P_2$ will cause an extra growth in $P_1$. This negative feedback, leads the system states to $P_{1max}$ and $P_{2min}$.

By using the simple mathematical model presented in [9] for gene regulation process, the differential equations of this genetic circuit can be written as:

$$\frac{dP_1}{dt} = s_1 \times \frac{k_{11}^{h_{11}}}{TF^{h_{11}} + k_{11}^{h_{11}}} \times \frac{k_{12}^{h_{12}}}{P_2^{h_{12}} + k_{12}^{h_{12}}} - d_1 P_1$$

$$\frac{dP_2}{dt} = s_2 \times \frac{k_2^{h_2}}{P_1^{h_2} + k_2^{h_2}} - d_2 P_2$$

(1)

Where $d_1$ and $d_2$ denote the degradation rate of $P_1$ and $P_2$. $s_1$ and $s_2$ represent maximum additional production rates arising from promoters activation. $k_{11}$ and $h_{11}$ are respectively Hill constant and Hill coefficient for binding of $TF$ repressor to promoter of the first gene. $k_{12}$ and $h_{12}$ are Hill constant and Hill coefficient for binding of $P_2$ repressor to promoter of the first gene. And $k_2$ and $H_2$ denote Hill constant and Hill coefficient for binding of $P_1$ repressor to promoter of the second gene.

**Numerical example:**

Figure 2 shows the steady state value of $P_2$ for different $TF$ values. As shown in this figure, at the steady state, if $TF < 1 \, mM$ then the value of $P_2$ will be about zero and if $TF > 1 \, mM$ then the value of $P_2$ will be around $2 \, mM$. Therefore, this input- output system has a switch-on relay like behavior.

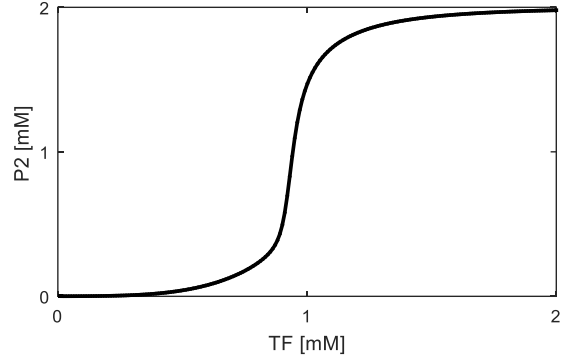

Figure 2: steady state value of $P_2$ with respect to $TF$. ($s_1 = 5 \times 10^{-6}\ M\ min^{-1}, s_2 = 10^{-6}\ M\ min^{-1}, k_{11} = 2 \times 10^{-7}\ M, k_{12} = 10^{-6}\ M, k_2 = 10^{-6}\ M, h_{11} = h_{12} = h_2 = 2, d_1 = 10^{-7}\ M\ min^{-1}, d_2 = 5 \times 10^{-7}\ M\ min^{-1}$)

*2.1. Setting switch-on point to an arbitrary value*

The switch-on point of the designed synthetic *Relay Switch* is on $T = 1\ mM$. But the switch-on point can be set to another value by changing the efficacy of the input ($TF$) on regulation of the first gene. This can be achieved by inhibition of $TF$. For example, if *Lac* operator represses the expression of the first gene, the efficacy of this transcription factor on gene expression can be controlled by Isopropyl β-d-1-thiogalactopyranoside (IPTG) [6]. Inhibition of *Lac* operator by *IPTG* can be described by [6]:

$$T = \frac{T_{tot}}{1 + k_I I^2} \qquad (2)$$

Where $TF_{tot}$ and $TF$ represent the whole concentration of Lac operator and effective value of Lac that acts as transcription factor for the first gene, respectively. The parameter $I$, denotes the concentration of IPTG. Note that $TF_{na} = TF_{tot} - TF$ is a representation of the inhibited Lac. Therefore, the required value of IPTG for setting the switch-on point to $TF = Sp$, is:

$$I = \sqrt{\frac{Sp - 1}{k_I}} \qquad (3)$$

**3. Synthetic comparator**

For feedback control of gene expression process, it is needed to compare the process output and the reference input. For this purpose, the design of synthetic genetic comparators is required, in order to compare the concentration of the two transcription factors. In this section, some synthetic comparators are presented, among which, the *Discrete Comparator* and the *Bistable Comparator* will be used for feedback control of gene expression process in section 4.

*3.1. Shifted Subtractor*

Since the concentration of a gene product has a positive value, design of a genetic subtractor is unreachable. While, a *Shifted Subtractor* can be used to compare concentration of two transcription factors. The *Shifted Subtractor* is described by following rule sets:

$$\text{If } TF_1 = TF_2 \text{ then } P_1 = \alpha$$

$$\text{If } TF_1 \to TF_{max} \text{ and } TF_2 \to 0 \text{ then } P_1 \to \alpha + \beta$$

$$\text{If } TF_1 \to 0 \text{ and } TF_2 \to TF_{max} \text{ then } P_1 \to \alpha - \beta$$

Where, $TF_1$ and $TF_2$ are the inputs and $P_1$ is the output of *Shifted Subtractor*. In addition, $\alpha$ and $\beta$ are positive values which $\alpha > \beta$.

**Proposition 2.** The genetic circuit depicted in Figure 3 acts as a genetic *Shifted Subtractor*.

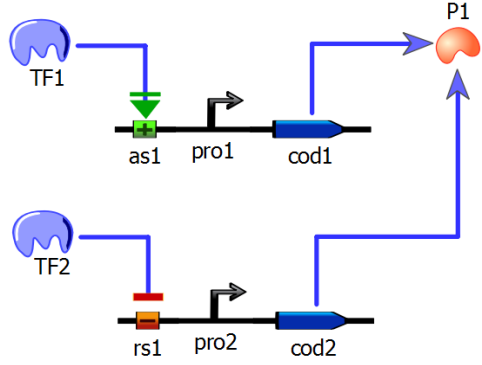

**Figure 3:** schematic of designed genetic *Shifted Subtractor* in TinkerCell software.

**Proof**: The differential equation of designed circuit is:

$$\frac{dP_1}{dt} = \frac{s_1 TF_1^{h_1}}{TF_1^{h_1} + k_1^{h_1}} + \frac{s_2 k_2^{h_2}}{TF_2^{h_2} + k_2^{h_2}} - d_1 P_1 \qquad (4)$$

Where $d_1$ denotes the degradation rate of $P_1$. $s_1$ and $s_2$ represent maximum additional production rates arising from promoters activation. $k_1$ and $h_1$ are respectively Hill constant and Hill coefficient for binding of $TF_1$ activator to promoter of the first gene. And $k_2$ and $h_2$ are respectively Hill constant and Hill coefficient for binding of $TF_2$ repressor to promoter of the second gene.

At the steady state:

$$P_1 = \frac{1}{d_1}\left(\frac{s_1 TF_1^{h_1}}{TF_1^{h_1} + k_1^{h_1}} + \frac{s_2 k_2^{h_2}}{TF_2^{h_2} + k_2^{h_2}}\right) \qquad (5)$$

If $s_1 = s_2 = s$, $h_1 = h_2 = h$ and $k_1 = k_2 = k$, then:

$$P_1 = \frac{s}{d_1}\left(\frac{(TF_1 TF_2)^h + 2(kTF_1)^h + k^{2h}}{(TF_1 TF_2)^h + (kTF_1)^h + (kTF_2)^h + k^{2h}}\right) =$$

$$\frac{s}{d_1}\left(1 + \frac{(kTF_1)^h - (kTF_2)^h}{(TF_1 TF_2)^h + (kTF_1)^h + (kTF_2)^h + k^{2h}}\right)$$

(6)

Accordingly

$$\alpha = \frac{s}{d_1}; \quad \beta = \frac{s \times k^h}{d_1}\left(\frac{TF_{max}^2}{TF_{max}^2 + k^h}\right); \tag{7}$$

**Numerical example:**

The numerical simulations implemented in MATLAB [10] depicted in Figure 4. Figure 4-A demonstrates the steady state plot of $P_1$ with respect to $TF_1$ and $TF_2$. Also, Figure 4-B shows the possible steady state values of $P_1$ corresponding to difference of $TF_1$ and $TF_2$. For example, for $TF_1 - TF_2 = 2mM$ the concentration of $P_1$ can be between $6.538 mM$ and $7.002 mM$. It's clear that the accurate value depends on the value of $TF_1$ and $TF_2$.

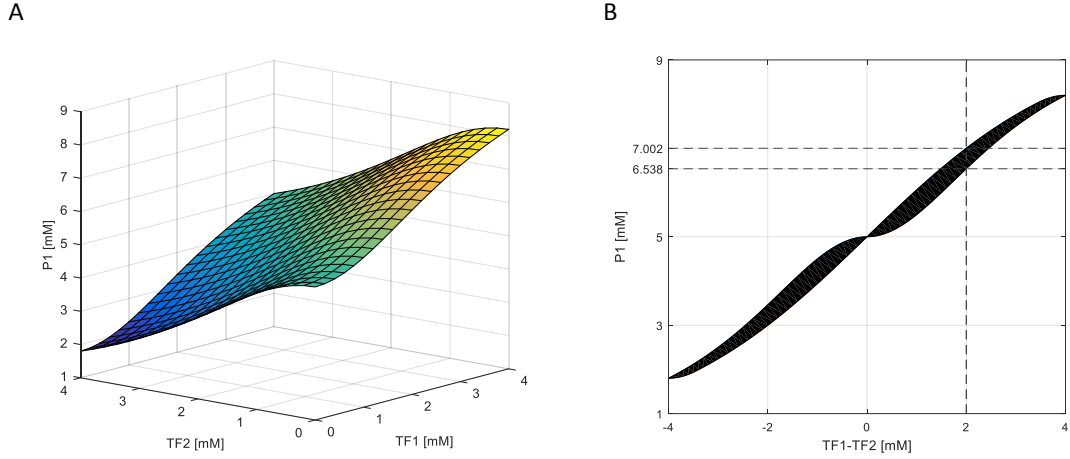

Figure 4: Numerical simulation of *Shifted Subtractor*. A: Steady state plot of $P_1$ with respect to $TF_1$ and $TF_2$. B: The possible steady state values of $P_1$ corresponding to difference of $TF_1$ and $TF_2$. ($s_1 = s_2 = 10^{-6}\ M\ min^{-1}, k_1 = k_2 = 3 \times 10^{-6}\ M, h_1 = h_2 = 2, d_1 = 2 \times 10^{-7}\ M\ min^{-1}$)

*3.2. Discrete Comparator*

While the concentration of a transcription factor has a positive value, the control signal takes a positive value in feedback control of gene expression process. Lack of negative control signal, necessitates the use of switching control method to control the gene expression process. This means that:

$$if\ P_{out} \geq T_{in}\ then\ P_c = P_{min}$$

$$if\ P_{out} < T_{in}\ then\ P_c = P_{max}$$

Which needs a *Discrete Comparator* as follows

$$if\ TF_1 \geq TF_2\ then\ P_c = P_{min}$$

$$if\ TF_1 < TF_2\ then\ P_c = P_{min}$$

**Proposition 3.** The genetic circuit depicted in Figure 5 acts as a genetic *Discrete Comparator*. $TF_1$ and $TF_2$ are the inputs and $P_3$ is considered as the output of the comparator.

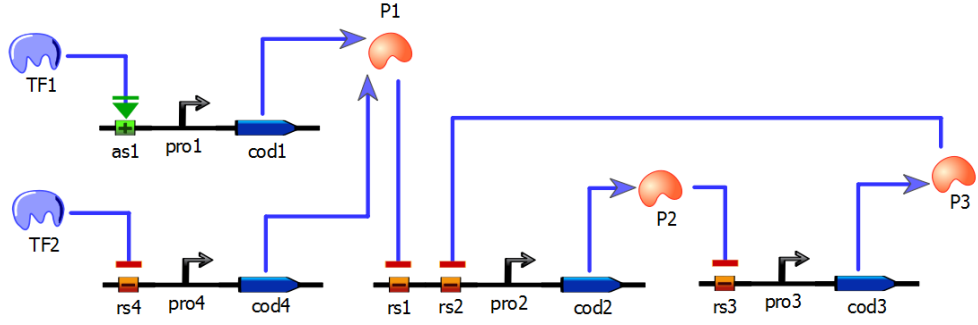

Figure 5: schematic of designed genetic *Discrete Comparator* in TinkerCell software.

**Proof**:

The designed *Shifted Subtractor* returns a value greater than α for $TF_1 > TF_2$ and less than α for $TF_1 < TF_2$. Addition of a *Relay Switch* to this comparator yields a genetic *Discrete Comparator* (Figure 5). It should be noted that the switch-on point should be set on α. The differential equations of this comparator are:

$$\frac{dP_1}{dt} = s_1 \frac{TF_1^{h_1}}{TF_1^{h_1} + k_1^{h_1}} + s_4 \frac{k_4^{h_4}}{TF_2^{h_4} + k_4^{h_4}} - d_1 P_1$$

$$\frac{dP_2}{dt} = s_2 \frac{k_{21}^{h_{21}}}{P_1^{h_{21}} + k_{21}^{h_{21}}} \times \frac{k_{22}^{h_{22}}}{P_3^{h_{22}} + k_{22}^{h_{22}}} - d_2 P_2 \qquad (8)$$

$$\frac{dP_3}{dt} = s_3 \frac{k_3^{h_3}}{P_2^{h_3} + k_3^{h_3}} - d_3 P_3$$

**Numerical example:**

Figure 6 illustrates the numerical simulation of designed *Discrete Comparator*. Figure 6-A demonstrates the steady state plot of $P_3$ with respect to $TF_1$ and $TF_2$. Also, Figure 6-B depicts the possible steady state values of $P_3$ with respect to difference of $TF_1$ and $TF_2$.

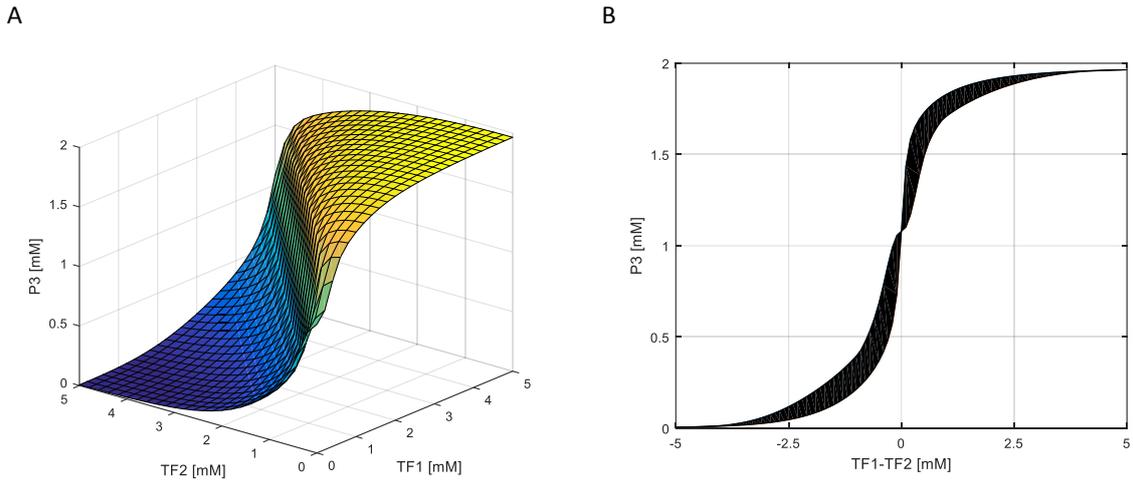

Figure 6: Numerical simulation of genetic *Discrete Comparator*. A: Steady state plot of $P_3$ with respect to $TF_1$ and $TF_2$. B: The possible steady state values of $P_3$ corresponding to difference of $TF_1$ and $TF_2$. ($s_1 = s_4 = s_3 = 10^{-6}\ M\ min^{-1}, s_2 = 5 \times 10^{-6}\ M\ min^{-1}, k_1 = k_4 = 3 \times 10^{-6}\ M, k_{21} = k_{22} = k_3 = 10^{-6}\ M, h_1 = h_{21} = h_{22} = h_3 = h_4 = 2, d_1 = 2 \times 10^{-7}\ M\ min^{-1}, d_2 = 10^{-7}\ M\ min^{-1}, d_3 = 5 \times 10^{-7}\ M\ min^{-1}$)

## 3.3. Bistable Comparator

**Proposition 3.** The genetic circuit depicted in Figure 7 acts as a genetic *Bistable Comparator*. This comparator constructed from two genes, where the protein expressed from each gene acts as the repressor for the other one. Moreover, each of them can be activated by an external inducer ($TF_1$ and $TF_2$). These inducers are considered as the circuit input and the comparison is made between them. Also, $P_1$ is chosen as the output of the comparator.

In this structure, the steady state value of protein concentration, corresponding to the activator with the greater concentration will take the maximum amount. Also the concentration of the produced protein of the gene with the lesser activator concentration will be minimum.

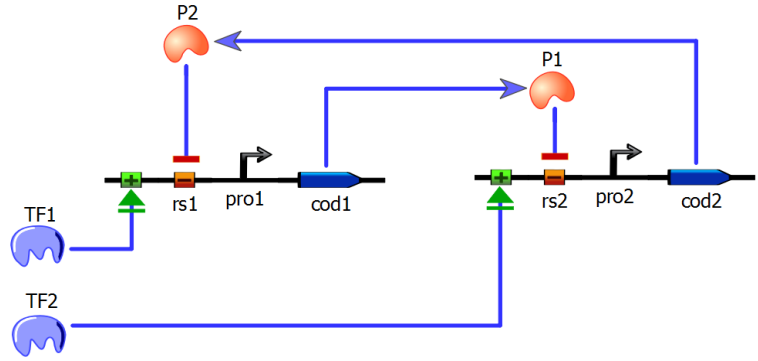

**Figure 7:** schematic of designed genetic *Bistable Comparator* in TinkerCell software.

**Proof:**

The differential equations of this system are:

$$\frac{dP_1}{dt} = \frac{s_1 TF_1^{h_{11}}}{TF_1^{h_{11}} + k_{11}^{h_{11}}} \times \frac{k_2^{h_{12}}}{P_2^{h_{12}} + k_{12}^{h_{12}}} - d_1 P_1$$
$$\frac{dP_2}{dt} = \frac{s_2 TF_2^{h_{21}}}{TF_2^{h_{21}} + k_{21}^{h_{21}}} \times \frac{k_{22}^{h_{22}}}{P_1^{h_{22}} + k_{22}^{h_{22}}} - d_2 P_2$$

(9)

Suppose that the both genes are similar and $P_1$ and $P_2$ have a same degradation rate. In other words, $s_1 = s_2$, $k_{11} = K_{21}$, $k_{12} = K_{22}$, $h_{11} = h_{12}$, $h_{12} = h_{22}$, $d_1 = d_2$. By this assumption, if $TF_1 > TF_2$ after a time, the concentration of $P_1$ and $P_2$ will rise and $P_1 > P_2$. According to this that $P_1$ and $P_2$ repress each other and $P_1 > P_2$, $P_2$ will have more degradation than $P_1$ and after a time we have $P_1 \gg P_2$. This negative feedback will go on and $P_1$ will take the maximum amount and $P_2$ will be nearly zero at the steady state. Also, if $TF_1 < TF_2$, at the steady state $P_1$ will be nearly zero and $P_2$ will take the maximum amount.

**Numerical example:**

Figure 8 demonstrates the numerical simulation of designed *Discrete Comparator*. Figure 8-A shows the steady state plot of $P_1$ with respect to $TF_1$ and $TF_2$ and Figure 8-B depicts the possible steady state values of $P_1$ corresponding to difference of $TF_1$ and $TF_2$.

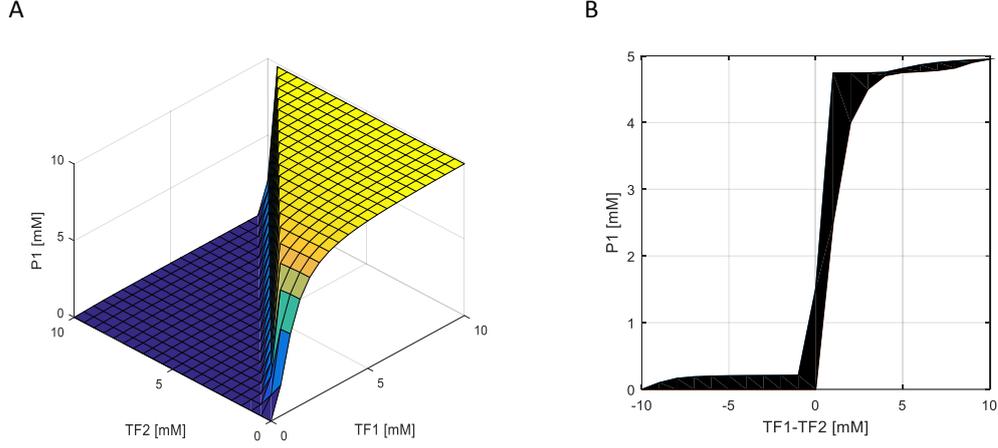

Figure 8: Numerical simulation of designed genetic *Bistable Comparator*. A: Steady state plot of $P_1$ with respect to $TF_1$ and $TF_2$. B: The possible steady state values of $P_1$ corresponding to difference of $TF_1$ and $TF_2$. ($s_1 = s_2 = 10^{-6}\ M\ min^{-1}, k_{11} = k_{12} = k_{21} = k_{22} = 10^{-6}\ M, h_{11} = h_{12} = h_{21} = h_{22} = 2, d_1 = d_1 = 10^{-7}\ M\ min^{-1}$).

## 4. Feedback control of gene expression process

Two different feedback controllers are provided depending on the type of the protein produced by the expression of process gene. In the case that the produced protein acts as an activator (type 1), the *Bistable Comparator* is employed to feedback control of gene expression process. But in the case that the produced protein acts as a repressor (type 2), a comparator consisting of a repressor input is required. And the *Discrete Comparator* is utilized in feedback control of the gene. These schemes are described in the following.

*4.1. Type 1*

If the protein produced from expression of process gene acts as an activator, the required comparators input must be an activator. Therefore, the *Bistable Comparator* is employed for feedback control of this kind of process gene (Figure 9). In this control system, the transcription factor $T_{in}$ is the input and the produced protein $P_{out}$ is the output of the system. Also, the protein $P_3$ acts as a control signal applied to the process. As the protein $P_1$ is in the type of repressor and the process gene is expressed by an activator transcription factor, the coding sequence of the first gene is composed of two regions for coding $P_1$ and $P_3$. $P_1$ is used to repress the expression of the second gene and $P_3$ is used to activate the process gene. The differential equations of the system are:

$$\frac{dP_1}{dt} = s_1 \frac{T_{in}^{h_{11}}}{T_{in}^{h_{11}} + k_{11}^{h_{11}}} \times \frac{k_{21}^{h_{21}}}{P_2^{h_{21}} + k_{21}^{h_{21}}} - d_1 P_1$$

$$\frac{dP_2}{dt} = s_2 \frac{P_{out}^{h_{21}}}{P_{out}^{h_{21}} + k_{21}^{h_{21}}} \times \frac{k_{22}^{h_{22}}}{P_1^{h_{22}} + k_{22}^{h_{22}}} - d_2 P_2$$

$$\frac{dP_3}{dt} = s_3 \frac{T_{in}^{h_{11}}}{T_{in}^{h_{11}} + k_{11}^{h_{11}}} \times \frac{k_{21}^{h_{21}}}{P_2^{h_{21}} + k_{21}^{h_{21}}} - d_3 P_3$$

$$\frac{dP_{out}}{dt} = s_4 \frac{P_3^{h_4}}{P_3^{h_4} + k_4^{h_4}} - d_4 P_{out}$$

(10)

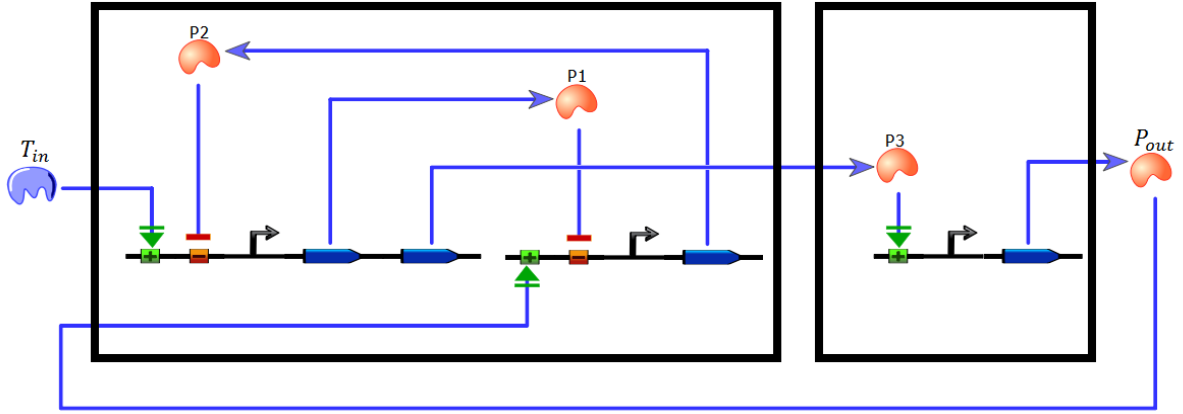

Figure 9: feedback control of gene expression process by *Bistable Comparator*.

In this system, the design parameters are $s_1, s_2, s_3, d_1, d_2, k_{12}, k_{22}, h_{12}$ and $h_{22}$, Since changing of the process properties such as the types of $P_3$, $P_{out}$ and $T_{in}$ are inaccessible.

*4.2. Type 2*

While the protein produced from expression of the process gene, acts as a repressor transcription factor, a comparator containing a repressor input is needed for feedback control of the process gene. Thus, the *Discrete Comparator* is used in the feedback control for these cases of process genes. Since the protein $P_1$ is assigned to be a negative feedback signal for the *Relay Switch*, this protein acts as a repressor and it cannot be used as an activator in the control of the process gene. Due to this fact, an extra coding sequence is added for expression and production of the protein $P_4$ to be used as a control signal (Figure 10). The differential equations of the controlled system are:

$$\frac{dP_1}{dt} = s_1 \frac{T_{in}^{h_1}}{T_{in}^{h_1} + k_1^{h_1}} + s_5 \frac{k_5^{h_5}}{P_{out}^{h_5} + k_5^{h_5}} - d_1 P_1$$

$$\frac{dP_2}{dt} = s_2 \frac{k_{21}^{h_{21}}}{P_1^{h_{21}} + k_{21}^{h_{21}}} \times \frac{k_{22}^{h_{22}}}{P_3^{h_{22}} + k_{22}^{h_{22}}} - d_2 P_2$$

$$\frac{dP_3}{dt} = s_3 \frac{k_3^{h_3}}{P_2^{h_3} + k_3^{h_3}} - d_3 P_3 \qquad (11)$$

$$\frac{dP_4}{dt} = s_4 \frac{k_3^{h_3}}{P_2^{h_3} + k_3^{h_3}} - d_4 P_4$$

$$\frac{dP_{out}}{dt} = s_6 \frac{P_4^{h_6}}{P_4^{h_6} + k_6^{h_6}} - d_6 P_{out}$$

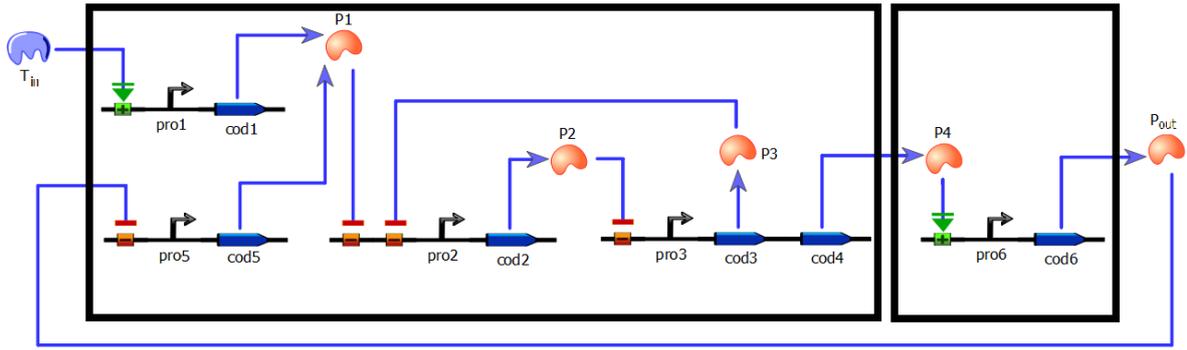

Figure 10: feedback control of the gene expression process by *Discrete Comparator*.

In this system, the design parameters are $\{s_1, s_2, s_3, s_4, s_5, k_{21}, k_{22}, k_3, h_{21}, h_{22}, h_3, d_1, d_2, d_3\}$.

## 5. Numerical results

The numerical simulation results of type 1 designed controller with differential equation 10 are represented in Figure 11. Description of parameters along with their values are summarized in table 1. Figure 11-A shows the steady state value of output $P_{out}$ (the concentration of protein produced by expression of process gene) for any values of input $T_{in}$. Also, the change of concentration in output $P_{out}$ over time for some values of input $T_{in}$ is shown in Figure 11-B.

**Table 1: Variables and parameters used in type 1 controller.**

| Parameter | Description | Value |
|---|---|---|
| Process parameters | | |
| $s_4$ | Max increase to $P_{out}$ production rate from activation (per plasmid) | $5 \times 10^{-6}$ M min$^{-1}$ |
| $k_4$ | Hill constant for binding of $P_3$ activator complex to promoter of target gene | $2 \times 10^{-6}$ M |
| $h_4$ | Hill coefficient for binding of $P_3$ activator complex to promoter of target gene | 2 |
| $d_4$ | Degradation rate of $P_{out}$ | $1 \times 10^{-7}$ M min$^{-1}$ |
| $k_{11}$ | Hill constant for binding of $T_{in}$ activator complex to promoter of Gene1 | $2 \times 10^{-6}$ M |
| $h_{11}$ | Hill coefficient for binding of $T_{in}$ activator complex to promoter of Gene1 | 2 |
| $k_{21}$ | Hill constant for binding of $P_{out}$ activator complex to promoter of Gene2 | $1 \times 10^{-6}$ M |
| $h_{21}$ | Hill coefficient for binding of $P_{out}$ activator complex to promoter of Gene2 | 2 |
| $d_3$ | Degradation rate of $P_3$ | $1 \times 10^{-6}$ M min$^{-1}$ |
| Design parameters | | |
| $s_1$ | Max increase to $P_1$ production rate from activation (per plasmid) | $3 \times 10^{-6}$ M min$^{-1}$ |
| $s_2$ | Max increase to $P_2$ production rate from activation (per plasmid) | $2 \times 10^{-6}$ M min$^{-1}$ |
| $s_3$ | Max increase to $P_3$ production rate from activation (per plasmid) | $3 \times 10^{-6}$ M min$^{-1}$ |
| $k_{12}$ | Hill constant for binding of $P_2$ repressor complex to promoter of Gene1 | $1 \times 10^{-6}$ M |
| $h_{12}$ | Hill coefficient for binding of $P_2$ repressor complex to promoter of Gene1 | 2 |
| $k_{22}$ | Hill constant for binding of $P_1$ activator complex to promoter of Gene2 | $2 \times 10^{-6}$ M |
| $h_{22}$ | Hill coefficient for binding of $P_1$ activator complex to promoter of Gene2 | 2 |
| $d_1$ | Degradation rate of $P_1$ | $1 \times 10^{-6}$ M min$^{-1}$ |
| $d_2$ | Degradation rate of $P_2$ | $1 \times 10^{-6}$ M min$^{-1}$ |

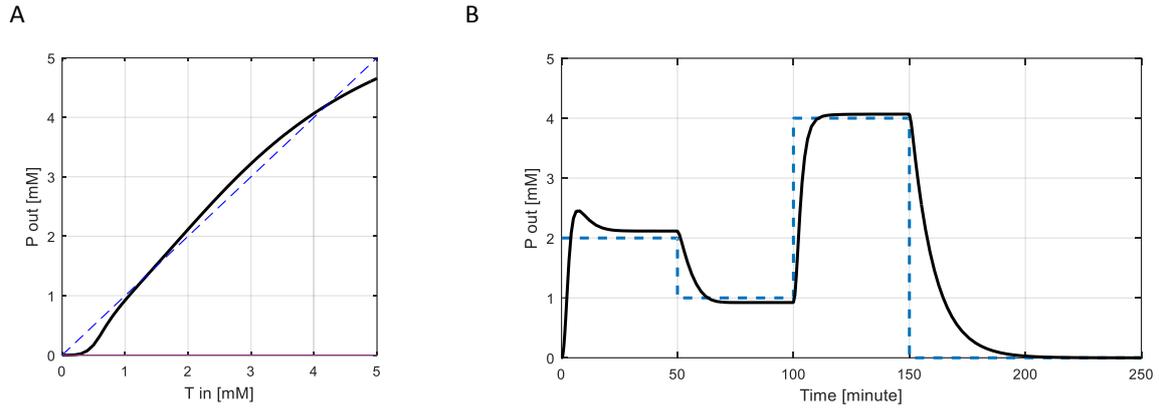

Figure 11: Setpoint regulation in the presence of type1 controller. (A) Steady state value of $p_{out}$ for any values of $T_{in}$. (B) Changing of $P_{out}$ over time for some values of $T_{in}$.

Simulation results for feedback control of type 2 gene expression process, described by differential equation 11 are shown in Figure 12. The parameters used in this simulation are presented in table 2. Figure 12-A shows the steady state value of output $P_{out}$ for different values of input $T_{in}$ and Figure 12-B shows the change of output $P_{out}$ over time for some values of input $T_{in}$.

**Table 2: Variables and parameters used in type 2 controller.**

| Parameter | Description | Value |
|---|---|---|
| Process parameters | | |
| $s_6$ | Max increase to $P_{out}$ production rate from activation (per plasmid) | $1 \times 10^{-6}$ M min$^{-1}$ |
| $k_6$ | Hill constant for binding of $P_4$ activator complex to promoter of target gene | $1 \times 10^{-6}$ M |
| $h_6$ | Hill coefficient for binding of $P_4$ activator complex to promoter of target gene | 2 |
| $d_6$ | Degradation rate of $P_{out}$ | $1 \times 10^{-7}$ M min$^{-1}$ |
| $k_1$ | Hill constant for binding of $T_{in}$ activator complex to promoter of Gene1 | $3 \times 10^{-6}$ M |
| $h_1$ | Hill coefficient for binding of $T_{in}$ activator complex to promoter of Gene1 | 2 |
| $k_5$ | Hill constant for binding of $P_{out}$ activator complex to promoter of Gene2 | $3 \times 10^{-6}$ M |
| $h_5$ | Hill coefficient for binding of $P_{out}$ activator complex to promoter of Gene2 | 2 |
| $d_4$ | Degradation rate of $P_4$ | $5 \times 10^{-7}$ M min$^{-1}$ |
| Design parameters | | |
| $s_1$ | Max increase to $P_1$ production rate from activation of Gene1 (per plasmid) | $1 \times 10^{-6}$ M min$^{-1}$ |
| $s_2$ | Max increase to $P_2$ production rate from activation (per plasmid) | $5 \times 10^{-6}$ M min$^{-1}$ |
| $s_3$ | Max increase to $P_3$ production rate from activation (per plasmid) | $1 \times 10^{-6}$ M min$^{-1}$ |
| $s_4$ | Max increase to $P_4$ production rate from activation (per plasmid) | $1 \times 10^{-6}$ M min$^{-1}$ |
| $s_5$ | Max increase to $P_1$ production rate from activation of Gene4 (per plasmid) | $1 \times 10^{-6}$ M min$^{-1}$ |
| $k_{21}$ | Hill constant for binding of $P_1$ repressor complex to promoter of Gene2 | $1 \times 10^{-6}$ M |
| $h_{21}$ | Hill coefficient for binding of $P_1$ repressor complex to promoter of Gene2 | 2 |
| $k_{22}$ | Hill constant for binding of $P_3$ repressor complex to promoter of Gene2 | $1 \times 10^{-6}$ M |
| $h_{22}$ | Hill coefficient for binding of $P_3$ repressor complex to promoter of Gene2 | 2 |
| $k_3$ | Hill constant for binding of $P_2$ repressor complex to promoter of Gene3 | $1 \times 10^{-6}$ M |
| $h_3$ | Hill coefficient for binding of $P_2$ repressor complex to promoter of Gene3 | 2 |
| $d_1$ | Degradation rate of $P_1$ | $2 \times 10^{-7}$ M min$^{-1}$ |
| $d_2$ | Degradation rate of $P_2$ | $1 \times 10^{-7}$ M min$^{-1}$ |
| $d_3$ | Degradation rate of $P_3$ | $5 \times 10^{-7}$ M min$^{-1}$ |

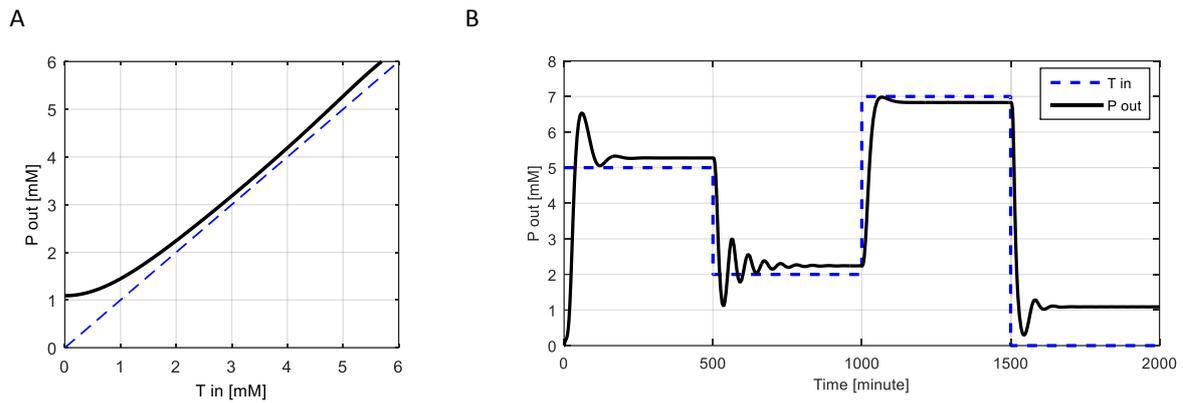

**Figure 12:** Setpoint regulation in the presence of type2 controller. (A) Steady state value of $p_{out}$ for any values of $T_{in}$. (B) Changing of $P_{out}$ over time for some values of $T_{in}$.

## 6. Conclusion

In this paper an intracellular feedback control method is proposed for tracking control of gene expression process. Firstly, Synthetic *Relay Switch* and *Shifted Subtractor* are proposed and are assembled in series to construct the *Discrete Comparator*. Also, *Bistable Comparator* have been designed for the first time in this paper. Then, the designed comparators are used to compare the process output with the desired input and tack specific control action. The numerical simulations demonstrate the effectiveness of the designed controllers.